\documentclass[aps, showkeys, reprint, onecolumn, superscriptaddress, notitlepage]{revtex4-1}

\usepackage[english]{babel}
\usepackage[utf8]{inputenc}
\usepackage{amsmath}
\usepackage{amsfonts}
\usepackage{amssymb}
\usepackage{graphicx}
\usepackage{array}
\usepackage{epsfig}

\newcommand{\diff}{\text{d}}

\usepackage{xcolor}

\begin{document}

\title{Excitation and resonant enhancement of axisymmetric internal wave modes}
\date{\today}

\author{S. Boury}
\affiliation{Univ Lyon, ENS de Lyon, Univ Claude Bernard, CNRS, Laboratoire de Physique, F-69342 Lyon, France}
\author{T. Peacock}
\affiliation{Department of Mechanical Engineering, Massachusetts Institute of Technology, Cambridge, MA 02139, USA}
\author{P. Odier}
\affiliation{Univ Lyon, ENS de Lyon, Univ Claude Bernard, CNRS, Laboratoire de Physique, F-69342 Lyon, France}

	\begin{abstract}
	
	To date, axisymmetric internal wave fields, which have relevance to atmospheric internal wave fields generated by storm cells and oceanic near-inertial wave fields generated by surface storms, have been experimentally realized using an oscillating sphere or torus as the source. Here, we use a novel wave generator configuration capable of exciting axisymmetric internal wave fields of arbitrary radial form to generate axisymmetric internal wave modes. After establishing the theoretical background for axisymmetric mode propagation, taking into account lateral and vertical confinement, and also accounting for the effects of weak viscosity, we experimentally generate and study modes of different order. We characterize the efficiency of the wave generator through careful measurement of the wave amplitude based upon group velocity arguments. This established, we investigate the ability of vertical confinement to induce resonance, identifying a series of experimental resonant peaks that agree well with theoretical predictions. In the vicinity of resonance, the wave fields undergo a transition to non-linear behaviour that is initiated on the central axis of the domain and proceeds to erode the wave field throughout the domain. 
	
	\end{abstract}
	\maketitle
	
	\section{Introduction}
	
	Since the early studies of G\"{o}rtler~\cite{gortler1943} and Mowbray \& Rarity~\cite{mowbray1967b}, laboratory experiments have played a central role in the development of understanding of internal wave fields. Initially, much of the focus was on two-dimensional internal wave beams generated by excitation methods such as an oscillating cylinder~\cite{mowbray1967b, sutherland1999} or moving topography~\cite{aguilar2006, echeverri2009}. For modeling purposes, such a wave field can be treated as nominally invariant in the transverse direction and thus described in terms of plane waves \textit{via} Fourier transforms~\cite{lighthill1967, sutherland2010}.
	
	Inspired by oceanographic considerations, and building on earlier experiments that utilized paddle generators to excite vertical~\cite{cacchione1974} or horizontal~\cite{thorpe1968a, thorpe1968b} modes, more recently, novel internal wave generator technology~\cite{gostiaux2006} has been utilized for a variety of studies of two-dimensional internal wave modes. In doubly-confined geometries (i.e. sidewalls, top and bottom), two-dimensional modes of different orders, determined by the combination of stratification, imposed frequency and dimensions of confinement, have been studied~\cite{benielli1998, sutherland2010}. The capability of the novel generator technology to investigate wave beams and two-dimensional modes was thoroughly explored by Mercier \textit{et al}~\cite{mercier2017}. Such capabilities have been employed to investigate, for example, the Triadic Resonant Instability (TRI) in a vertical mode propagating horizontally~\cite{joubaud2012} or the formation of multilayered stratifications~\cite{dossmann2017}. While theoretical studies for linear stratifications describe such wave fields in terms of the natural modal basis of sines and cosines, it should be recalled that the modal pattern can be considered as a combination of plane waves propagating and reflecting from the system boundaries~\cite{mercier2017}.
	
	Axisymmetric wave fields have traditionally been experimentally excited using an oscillating sphere and exploring the shape of the wave beams~\cite{ermanyuk2011, flynn2003, ghaemsaidi2013, mowbray1967a, peacock2005, stevenson1969}. While the form of the wave field close to the oscillating body is nontrivial, modeling studies have explored the limit states of the wave beams in terms of plane waves with a spherical amplitude decreasing as $r^{-1/2}$, $r$ being the radial distance from the sphere, computed from the Green function of the moving source~\cite{voisin2003}, or as infinite sums of Bessel functions with complex coefficients~\cite{davis2010, sutherland2010}.  The amplitude decrease and the viscous decay of the conical wave beam emitted by an oscillating sphere has been explored in laboratory experiments by Flynn \textit{et al}~\cite{flynn2003} showing good agreement with theoretical predictions. More sophisticated axisymmetric experimental geometries have been investigated using a vertically oscillating torus~\cite{duranmatute2013, ermanyuk2017}, in which case a highly non-linear process occurs due to the three-dimensional geometric focusing, able to transport momentum and break into turbulence. None of these experimental configurations, however, readily permitted a change in the form nor the wave number of the wave field being excited.
	
	An axisymmetric wave generator~\cite{maurer2017}, adapted from its planar counterpart~\cite{gostiaux2006}, has been demonstrated as capable of generating high-fidelity axisymmetric internal wave fields, with substantial flexibility in the setting of the radial wavelength. Studies using this technology reveal axisymmetric wave cones propagating in the stratified medium according to the internal wave dispersion relation and with radial profiles imposed by the configuration of the generator, such as ring-shaped excitation or truncated Bessel functions~\cite{maurer2017}. Wave amplitudes and frequencies were measured, showing a good agreement with the linear theory for axisymmetric waves in a stratified fluid of constant buoyancy, both in the non-rotating and rotating cases. 
	
	To date, there have been no experimental studies of internal wave modes in an axisymmetric geometry. Furthermore, to our knowledge, there is no quantitative study of resonant confined modes, even in two-dimensional geometries. In this paper, we perform laboratory experimental realizations of axisymmetric modes, made possible by the new form of wave generator developed by Maurer \textit{et al}~\cite{maurer2017}. In section $2$ we establish the general theory for axisymmetric modes of internal waves by considering both radial and vertical confinement as well as weakly viscous effects. Then, in section $3$, we describe our experimental apparatus, adapted from Maurer's~\cite{maurer2017}. Experimental results are presented in section $4$, followed by conclusions and discussion in section $5$.

	\section{Theory}
	
		\subsection{Governing Equations}
		
		In a cylindrical framework ($\mathbf{e_r}$, $\mathbf{e_\theta}$, $\mathbf{e_z}$), with $\mathbf{e_z}$ the ascendent vertical, small amplitude inertia gravity waves in an inviscid rotating fluid with a constant background stratification satisfy the following equations in the Boussinesq approximation:
		\begin{align}
			\rho_0 \left(\dfrac{\partial \mathbf{v}}{\partial t} + \left(\mathbf{v}\cdot\mathbf{\nabla}\right) \mathbf{v}\right) &= - \rho_0 f \mathbf{e_z} \times \mathbf{v} - \mathbf{\nabla} p - (\rho - \bar{\rho}) g \mathbf{e_z},\label{eq:IW1}\\
			\dfrac{\partial \rho}{\partial t} + \left(\mathbf{v}\cdot \mathbf{\nabla}\right) \rho &= 0,\label{eq:IW2}\\
			\mathbf{\nabla} \cdot \mathbf{v} &= 0\label{eq:IW3},
		\end{align}
		where $\mathbf{v} = (v_r,~v_\theta,~v_z)$ is the velocity field, $p$ the pressure field, $\rho$ the density field, and $\bar{\rho}$ the background density field. We define the Coriolis frequency $f= 2 \Omega$ as twice the rotation frequency $\Omega$, and the buoyancy frequency $N$ \textit{via} the relation $N^2 = (-g/\rho_0) \partial \bar{\rho} / \partial z$ with $\rho_0$ being a reference density.
		
		Considering axisymmetric wave fields, we assume that there is no dependence in the orthoradial variable $\theta$ and hence all functions only depend on ($r$, $z$, $t$). By introducing the axisymmetric stream function $\psi$ such that:
		\begin{equation}
			v_r = - \frac{1}{r}\dfrac{\partial (r \psi)}{\partial z} \mathrm{~~~~~~~and~~~~~~~} v_z =\frac{1}{r}\dfrac{\partial (r \psi)}{\partial r},\label{eq:IW4}
		\end{equation}
		equations~\eqref{eq:IW1},~\eqref{eq:IW2}, and~\eqref{eq:IW3} become:
		\begin{align}
			\partial_t^2 \left(\partial_z^2 \psi + \partial_r \left(\frac{1}{r}\partial_r (r \psi) \right) \right) + N^2 \partial_r \left(\frac{1}{r}\partial_r (r\psi)\right) + f^2 \partial_z^2 \psi = 0.\label{eq:IW5}
		\end{align}
Natural axisymmetric solutions of this equation can be found through a Fourier-Hankel decomposition. Using a modal basis, the solutions write as linear combinations of Bessel functions of the first kind $J_1$ and of the second kind $Y_1$. They all lead to the same theoretical study, but $Y_1$ has a singularity at $r=0$ so only the $J_1$ function will be considered. As different radial wave numbers may enter in the decomposition, the stream function $\psi$ can then be written as a modal sum:
		\begin{equation}
			\psi (r,z,t) = \iint \phi (z) J_1 (l r) \exp (-i \omega t) \diff l \diff \omega.\label{eq:IW6}
		\end{equation}
		
		Radial and vertical velocities can be derived from equation~\eqref{eq:IW6} using classic relations for the Bessel derivatives, as follows:
		\begin{align}
			v_r &= \iint \phi'(z) J_1(lr) \exp(-i \omega t) \diff l \diff \omega, \label{eq:IW7}\\
			v_z &= \iint l \phi(z) J_0(lr) \exp(-i \omega t) \diff l \diff \omega.\label{eq:IW8}
		\end{align}
For a given frequency $\omega$ and radial mode $l$, $\phi(z)$ satisfies:
		\begin{equation}
			(f^2 - \omega^2) \phi'' (z) -l^2 (N^2 - \omega^2) \phi (z) = 0.\label{eq:IW9}
		\end{equation}
Solutions of equation~\eqref{eq:IW9} are exponential functions, either complex or real. They can be either propagative or evanescent waves, depending on the frequency, as long as the vertical wave number $m$ satisfies the dispersion relation:
		\begin{equation}
			m^2 =  l^2 \frac{N^2 - \omega^2}{\omega^2 - f^2}.\label{eq:IW10}
		\end{equation}
		
		The effect of rotation on a conical wave beam emitted in a stratified fluid has already been documented in previous studies~\cite{maurer2017, peacock2005}. Hence, results for gravity waves could be easily extended to a rotating case with inertial or, more generally, inertia-gravity waves, as it mainly changes the vertical wave number $m$. In this study, we focus on a non-rotating fluid ($f=0$). The dispersion relation simplifies and if we define $\beta = \sin ^{-1} (l/k)$ to be the angle between the vertical axis and the wave vector $\mathbf{k}=(l,0,m)$, we obtain:
		\begin{equation}
			\sin \beta = \pm \frac{\omega}{N}.\label{eq:IW11}
		\end{equation}
According to equation~\eqref{eq:IW11}, internal gravity waves propagate along a direction fixed by the angle $\beta$. In a two dimensional geometry, four wave beams on a St Andrew's Cross are formed~\cite{mowbray1967b, sutherland1999}. In a three dimensional axisymmetric geometry, the dispersion relation sets two cones aligned along the vertical direction and connected by the apex~\cite{peacock2005, sutherland2010}.

		\subsection{Radial Confinement}
		
\begin{figure}[!htbp]
				\centering
				\epsfig{file=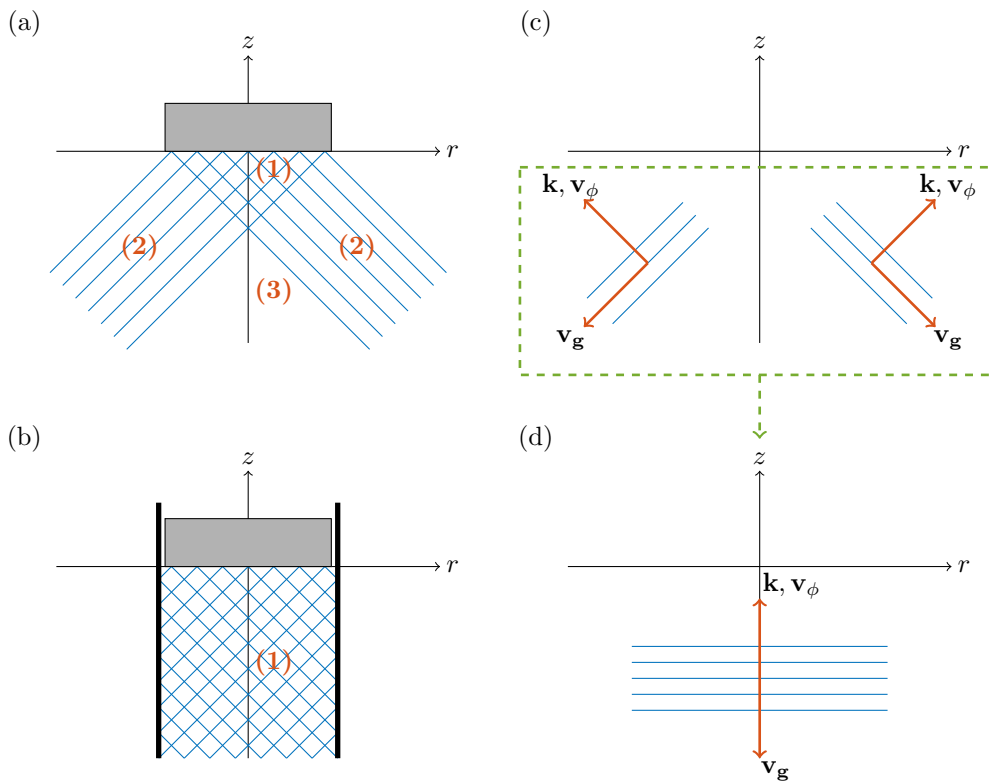} 
				\caption{Left: beams for downwards propagating wave generation in an (a) unconfined and (b) confined geometries produced by a wave generator or an oscillating body (light grey rectangle). Right: (c) phase lines of two different wave beams allowed by the dispersion relation and (d) phase lines of the wave re-recombination for vertically propagating horizontal (radial) modes.}
				\label{fig:1}
			\end{figure}

		In a previous study, Maurer \textit{et al}~\cite{maurer2017} analysed the production of a conical wave field generated by an axisymmetric moving form at the surface, for which the radial profile was a truncated Bessel function. Although Bessel functions form a natural basis of study for axisymmetric wave fields, the analytical form of the wave field for a truncated Bessel function is not so simple. An illustration is presented in figure~\ref{fig:1}(a), which displays a vertical cut of the spatial structure of the wave field studied in~\cite{maurer2017}. Immediately below the generator (region $1$ in figure~\ref{fig:1}(a)), the wave field preserves its radial form, but further below the wave field develops a conical beam-shaped profile (region $2$), which can locally be modelled by a plane wave. Finally, due to the propagation angle set by~\eqref{eq:IW11}, sufficiently far below the oscillating body around the vertical axis the wave field is absent (region $3$). Analytically, this evolution of the wave field is a natural consequence of the truncated Bessel function forcing being expressed as an integral over Bessel functions of different wavelengths, with coefficients depending on the spatial forcing. Complex models have been set up to understand the nature of such radiated wave fields~\cite{davis2010, flynn2003, voisin2011}.
			
		Given the finite spatial extent of the forcing, to generate a wave field of simple form expressible using a single radial Bessel function, confinement can be imposed experimentally to the fluid, as illustrated in figure~\ref{fig:1}(b). As for planar geometries, confinement prevents the wave from propagating in the bounded direction. Given the assumption of axisymmetry, we seek a wave field that is radially confined by a cylinder of radius $R$ equal to the radius of the generator, and vertically propagating, which is in contrast to the planar scenario that has vertical confinement and permits lateral propagation~\cite{dossmann2017, joubaud2012, mercier2017}. We impose the radial boundary condition:
		\begin{equation}
			v_r (r=R,z) = \left( \dfrac{\partial\psi}{\partial z}\right) _{(r=R, z)} = 0, \label{eq:IW12}
		\end{equation}
which corresponds to a maximum of vertical velocity and a vanishing radial velocity at the outer boundary, and conserves volume in the domain. Condition~\eqref{eq:IW12} limits possible values of the radial wave number $l$ and if the fluid is excited with this wavenumber at  frequency $\omega$, a single propagating mode is expected to result.
		
			Figure~\ref{fig:1}(c) shows the two directions of propagation allowed for the wave beams by the dispersion relation~\eqref{eq:IW11}, in a planar cross-section, for a wave generation at the surface. In our experiment, the radial confinement leads to downward propagating modes which are, as depicted in figure~\ref{fig:1}(c,d) (still in a planar cross-section) a re-combination of conically propagating beams. Due to the symmetry with respect to the vertical axis, the radial direction of propagation cancels out and, for a downwards propagating wave at a selected frequency $\omega$ and wavenumber $\mathbf{k}$, the phase and group velocities can be computed from the dispersion relation~\eqref{eq:IW11}:
		\begin{align}
			\mathbf{v_\phi} &= \frac{1}{2\pi} \int_0 ^{2\pi} \frac{N l}{k}(l \mathbf{e_r} + m \mathbf{e_z}) \diff \theta = \frac{N l m}{k}\mathbf{e_z},
			\label{eq:IW14a}\\
			\mathbf{v_g} &=  \frac{1}{2\pi} \int _0 ^{2\pi} \frac{m l N^2}{\omega k^4} (m\mathbf{e_r}-l\mathbf{e_z}) \diff \theta = - \frac{m l^2 N^2}{\omega k^4} \mathbf{e_z}.
			\label{eq:IW14b}
		\end{align}
		Equations~\eqref{eq:IW14a} and~\eqref{eq:IW14b} show phase and group velocity oriented in opposite directions, illustrated in figure~\ref{fig:1}(d), consistent with oceanic signatures identified by oceanographers looking for internal waves~\cite{alexander1995, alford2001}. This feature contrasts with horizontally propagating cartesian modes which show phase and group velocities pointing towards the same direction~\cite{mercier2017}.
		
			To investigate the shape of the wave field in the experimental domain, and more specifically its amplitude, we extend the axisymmetric analysis of Sutherland~\cite{sutherland2010}, first derived for an oscillating cylinder in a two-dimensional geometry, by applying it to our axisymmetric flat generator in a confined domain. Neglecting rotation, equation~\eqref{eq:IW5} can be rewritten:
			\begin{equation}
				\Gamma^2 \dfrac{\partial}{\partial r}\left(\frac{1}{r}\dfrac{\partial (r \psi)}{\partial r} \right) + \dfrac{\partial^2 \psi}{\partial z^2} = 0,\label{eq:IW15}
			\end{equation}
			with $\Gamma^2 = 1 -N^2/\omega^2$. Through a Fourier transform, the time dependency of the streamfunction can be expressed in complex coordinates as $\psi \propto e^{-i \omega t}$, the velocity field being the real part of the stream function derivatives.
			
			Considering that the plates of the generator are moving vertically and are injecting a vertical velocity $a \omega$, with $a$ being a sufficiently small displacement so that the fluid surface can be considered to reside at $z=0$, the boundary conditions that apply to the streamfunction field are:
			\begin{align}
				v_z (r, z=0) = \left(\frac{1}{r}\dfrac{\partial (r \psi)}{\partial r}\right)_{(r, z=0)} &= a \omega J_0 (l r), \label{eq:IW16}\\
				v_r (r=R,z) = \left( \dfrac{\partial\psi}{\partial z}\right) _{(r=R, z)} &= 0.\label{eq:IW17}
			\end{align}
			Equation~\eqref{eq:IW16} means that the generator imposes its own movement to the fluid at the top of the domain. The modal boundary condition is expressed by equation~\eqref{eq:IW17} as detailed before.
			
			For $N<\omega$, the problem can be readily solved \textit{via} a coordinate transformation: $(r' = \Gamma r,~z'=z)$ so that equation~\eqref{eq:IW15} becomes:
			\begin{equation}
				\Delta_h'\psi = 0,\label{eq:IW21}
			\end{equation}
			where $\Delta'_h$ is the horizontal Laplacian. The solution can be obtained using separation of variables. The radial part of the equation satisfies a Bessel differential equation of first order, leading to $\psi \propto J_1(l r)$. The vertical component is found to be exponential (see equation~\eqref{eq:IW9}), and $\psi \propto \exp(\Gamma l z)$ as the amplitude decreases as $z$ goes to $-\infty$. From the boundary conditions, the different coefficients can be set and recasting the solution in the original coordinates, we obtain:
			\begin{equation}
				\psi_{N<\omega} (r,z,t) = -\frac{a \omega}{l} J_1 (l r) \exp (m z) \cos (\omega t),\label{eq:IW22}
			\end{equation}
			where we define $m = \Gamma l$ as the vertical wave number, which contains the influence of the stratification.
			
			In the case $N > \omega$, the term $1 - N^2/ \omega^2$ is negative. We thus define $\gamma^2 = \omega^2/N^2 - 1$ and, by analytic continuation, the problem can be solved using the same method as before. Thanks to the second order derivatives, the problem keeps well-defined though we are using complex analysis and the final stream function belongs to the real space of functions:
			\begin{equation}
				\psi_{N>\omega} (r,z,t) = -\frac{a \omega}{l} J_1 (l r) \cos (m z - \omega t),\label{eq:IW23}
			\end{equation}
			with $m$ defined as $m^2 = - \gamma^2 l^2$. Hence, we obtain two different radial modes, one being evanescent~\eqref{eq:IW22} and the other one propagating in the vertical direction~\eqref{eq:IW23}.
			
			The vertical velocity being a radial derivative, it behaves as $v_z \propto a \omega$:
			\begin{align}
				v_z (r,z) = \frac{1}{r}\dfrac{\partial (r \psi)}{\partial r} &= a \omega J_0 (l r) \cos(mz -\omega t), \label{eq:IW40}\\
				v_r (r,z) = \dfrac{\partial\psi}{\partial z} &= -\frac{a \omega m}{l} J_1(l r) \sin(m z -\omega t).\label{eq:IW41}
			\end{align}
					
		\subsection{Vertical Confinement}
			
			Due to the finite extent of the experimental domain, boundaries at the top ($z=0$) and at the bottom ($z=-L$) are to be taken into account. This creates a cavity that behaves like an opto-electromagnetic cavity~\cite{jackson1999} or a Melde's string~\cite{melde1860}, with different modes and mechanical resonances. The total wave field in the cavity is obtained by a superposition of all the reflected waves, from the top and the bottom of the tank, creating constructive or destructive interferences. In this configuration, the generator is continuously exciting a velocity field of stream function $\psi_1$, written using the complex notation as:
			\begin{equation}
				\psi_1 (r, z, t) = \psi^0_1 J_1 (l r) e^{i(\omega t - m z)},\label{eq:IW27}
			\end{equation}
			with $\psi^0 = a \omega / l$. At $z=-L$, the downwards wave field $\psi_1$ is reflected into an upwards wave field $\psi_2$, and at $z=0$, the $\psi_2$ stream function is reflected into another downwards wave $\psi_3$. Repeated reflections occur at $z=0$ and $z=-L$, and as a result the total stream field is composed of an infinite sum of reflected wave fields.
	
			We denote by odd numbers the downwards waves and by even numbers the upwards waves. At the boundaries, as well as changing direction, reflection also induces a $\pi$ phase shift, and if we assume that there is no dissipation the amplitudes of the stream functions are equal before and after reflection. Boundary conditions at the top and at the bottom of the tank then apply as:
			\begin{align}
				\psi_{2k - 1} (z=-L) &= \psi_{2k} (z=-L) e^{i \pi}, \label{eq:IW28}\\
				\psi_{2k + 1} (z= 0) &= \psi_{2k} (z= 0) e^{i \pi}, \label{eq:IW29}
			\end{align}
			leading to:
			\begin{align}
				\psi^0_{2k} &= \psi^0_{2k - 1} e^{-2 i m L- i \pi}, \label{eq:IW30}\\
				\psi^0_{2k} &= \psi^0_{2k + 1} e^{i\pi}. \label{eq:IW31}
			\end{align}
We deduce that the general expression of these wave amplitudes are:
			\begin{align}
				\psi^0_{2k} &=  \psi^0_1 e^{-2ikmL + (2k-1)\pi}, \label{eq:IW32}\\
				\psi^0_{2k+1} &= \psi^0_1 e^{-2ikmL + (2k)\pi}. \label{eq:IW33}
			\end{align}
As the tank is filled by infinite wave reflections, we describe the total wave field by a sum over all the reflected waves:
			\begin{equation}
				\psi = \sum_{k=1}^\infty \psi_k = \psi^0_1 J_1 (l r) e^{i\pi / 2} e^{i \omega t} \frac{\sin (m(z - L))}{i \sin(mL)}, \label{eq:IW34}
			\end{equation}
			hence the real field becomes:
			\begin{equation}
				\Re (\psi) = \psi^0_1 J_1 (l r) \frac{\cos (\omega t) \sin (m(z-L))}{\sin (mL)}. \label{eq:IW35}
			\end{equation}
		
			Waves that contribute to the total wave field interact either constructively or destructively. In the first case, we would be able to define a temporal and a spatial period, fixed by the wave parameters $\omega$, $l$, and $m$, and by the size of the cavity $L$, as in any wave resonator. Exact cavity modes are obtained if the reflection at $z=-L$ produces a reflected wave in phase with the incoming wave, which means that this position is already a node of the wave field. This resonance condition can be expressed as:
			\begin{equation}
				L = n \frac{\lambda}{2}, \mathrm{~for~}n\in \mathbb{N},\label{eq:IW36}
			\end{equation}
			with $\lambda = 2\pi / m$ being the vertical spatial period. A direct consequence is that the reflection at $z=0$ also produces a wave in phase with the incoming wave, so all reflected waves will be interacting constructively. This relation can be written as a condition involving resonant frequencies $\omega_n$:
			\begin{equation}
				\frac{\omega_n}{N} = \frac{(L l)^2}{\pi^2 n^2 + (L l)^2}. \label{eq:IW39}
			\end{equation}
Similar to electromagnetic waves, the cavity operates as a frequency selector, as a discrete number of frequencies $\omega_n$ fulfills the resonance condition. We present in table~\ref{tab:resonantomegas} a list of the first ten resonant frequencies that can be selected in a radial mode$-1$ configuration with $L=60\mathrm{~cm}$ and $l=19\mathrm{~m^{-1}}$.
			\begin{table}[!htbp]
			\begin{center}
				\begin{tabular}{c c c c c c c c c c c c}
					\hline 
					$n$ & $0$ & $1$ & $2$ & $3$ & $4$ & $5$ & $6$ & $7$ & $8$ & $9$ & $10$ \\ 
					\hline \hline
					$\omega_n / N$ & $1$ & $0,964$ & $0,876$ & $0,771$ & $0,672$ & $0,588$ & $0,518$ & $0,460$ & $0,413$ & $0,374$ & $0,341$ \\ 
					\hline
				\end{tabular}
				\caption{First ten resonant frequencies computed for $L=60\mathrm{~cm}$ and $l=19\mathrm{~m^{-1}}$.\label{tab:resonantomegas}}
			\end{center}
			\end{table}
			
			\subsection{Weakly Viscous Correction}
			
			In the preceding derivations, an inviscid fluid was assumed. This made possible the propagation of a single mode at all frequencies without damping effect, and the existence of exact resonant cavity modes. As we will see, however, such an approximation is only relevant for a selected range of frequencies. To quantify the viscous effects the wave propagation, we write the vertical wave number as the following expansion:
			\begin{equation}
				m = m_0 + i \varepsilon m_1 + \mathcal{O}(\varepsilon^2),\label{eq:IW24}
			\end{equation}
			with $\varepsilon = \nu l^2 / \omega$, $m_0$ being the inviscid wave number (equation~\eqref{eq:IW10}), and $m_1$ being the first order correction. Including viscous terms, equation~\eqref{eq:IW9} becomes:
			\begin{equation}
				\phi^{(4)}(z) - \left(2 l^2 + i \frac{f^2-\omega^2}{\nu \omega}\right) \phi ''(z) + l^2 \left(l^2 + i \frac{N^2 - \omega^2}{\nu \omega} \right)\phi(z) = 0. \label{eq:IW25}
			\end{equation}
			Hence, with the vertical dependence being $\exp(i m z)$ (complex notation of equation~\eqref{eq:IW23}) and $m$ defined as in equation~\eqref{eq:IW24}, one can extract from equation~\eqref{eq:IW25} the following weakly viscous correction:
			\begin{equation}
				i \varepsilon m_1 = \mp \frac{i \varepsilon l}{2 (1 - \gamma^2) \alpha^3\sqrt{1-\alpha^2}},
			\end{equation}
			where $\alpha = \omega /N$ and $\gamma = f/\omega$. For a non-rotating case, this correction simplifies to:
			\begin{equation}
				i \varepsilon m_1 = \mp \frac{i \varepsilon l }{2 \alpha^3\sqrt{1-\alpha^2}}.
			\end{equation}
			At an altitude $z$ below the wave generation source, the weakly viscous streamfunction $\psi_\nu$ writes:
			\begin{equation}
				\psi_\nu (z) = \psi (z) \exp (-\varepsilon m_1 |z|).\label{eq:IW26}
			\end{equation}
			
			According to equation~\eqref{eq:IW26}, the typical vertical length of viscous damping $1/\varepsilon m_1$ depends on the frequency $\omega$ and is much smaller at low frequencies than at high frequencies. Quantifying theoretically the viscous damping will help us to understand amplitudes observed in the tank when we compare them to the velocity amplitudes of the generator.
			
	\section{Experimental Apparatus} 
	
\begin{figure}[!htbp]
			\centering
			\epsfig{file=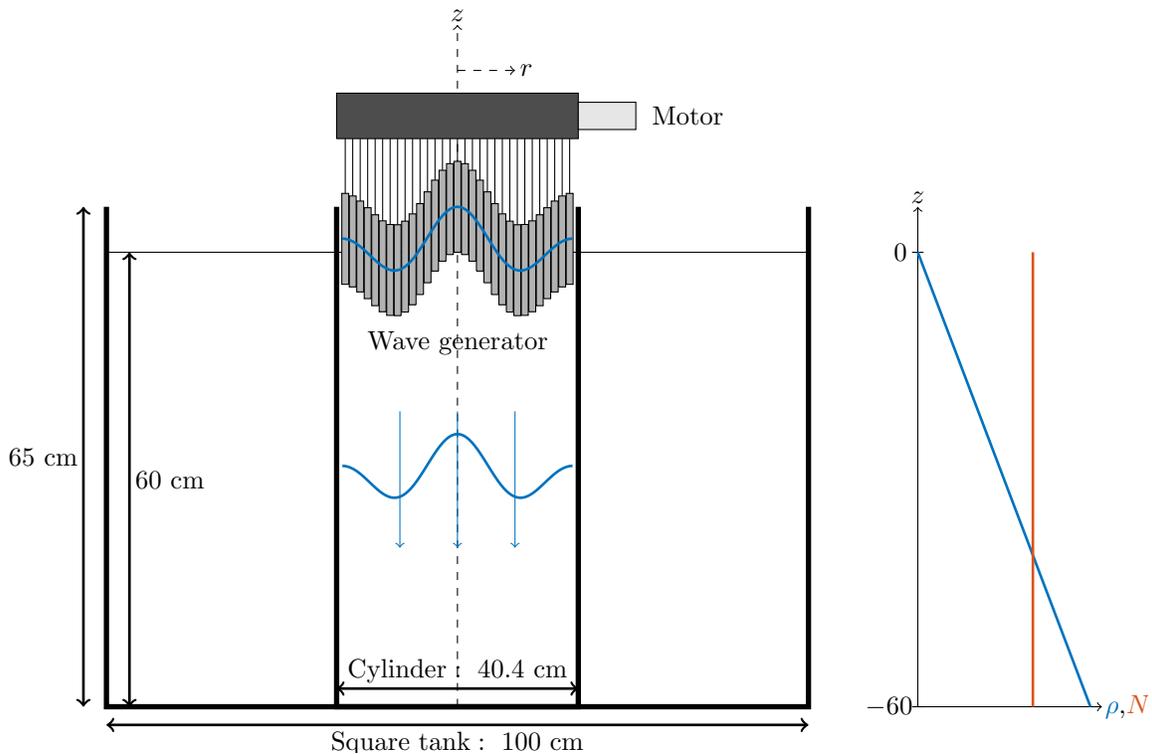}
			\caption{Schematic of the experimental apparatus. Left: a cylindrical tank, inside a square tank, confines the waves produced by the generator located at the surface, leading to a radial Bessel mode propagating downwards. Right: idealised linear stratification and constant buoyancy frequency.}
			\label{fig:fig2}
		\end{figure}

		To conduct our experiments, the experimental setup of Maurer \textit{et al}~\cite{maurer2017} was adapted. A general schematic of the experimental apparatus is presented in figure~\ref{fig:fig2}. We introduce natural cylindric coordinates with $z$, the ascendent vertical at the center of the apparatus, and $r$ the radius orthogonal to the vertical axis. The origin in $z$ is taken at the surface of the water.
		
		The generator comprises sixteen, $1.2\mathrm{~cm}$ wide, concentric PVC cylinders periodically oscillating, each of them being forced by two eccentric cams. The eccentricities can be configured to introduce a phase shift between the different cylinders, and the oscillating amplitude can be set for each individual cylinder. As a result, the vertical displacement of the $n^{th}$ cylinder can be described by:
		\begin{equation}
			a_n(t) = A_n \cos (\omega t + \theta),
		\end{equation}
		with $A_n$ its amplitude, $\omega$ the forcing frequency, and $\theta$ a phase shift. For a smooth motion of the PVC cylinders, a $0.1\mathrm{~cm}$ gap is kept between each cylinder and the total diameter of the wave generator is then $40.2\mathrm{~cm}$. The generator is mounted at the surface of the water to force downwards internal waves.
		
				To investigate the ability of this experimental setup to produce modal wave fields, we set the generator in three different configurations to excite first, second, and third order modes. Modes are defined by the number of nodes of the Bessel function present in the range $r \in [0;~20]\mathrm{~cm}$ (size of the generator). It sets the radial wavelength and we computed the three associated wave numbers: $l_1=19\mathrm{~m^{-1}}$, $l_2=35\mathrm{~m^{-1}}$, and $l_3=51\mathrm{~m^{-1}}$. We did not look for modes of higher orders because the discretisation of the generator profile was not sufficient to produce smooth enough shapes of Bessel functions. The amplitudes of the different cams for the three modes are summarized in table~\ref{tab:shafts}. The profiles can be defined by the radial wavenumber $l$ and the amplitude at $r=0$ that we call $a$. The different amplitudes $a_n$, for $n\neq 0$, are taken to be the discrete approximation of the Bessel function defined by $l$ and $a$.	
			\begin{table}[!htbp]
				\begin{center}
					\begin{tabular}{c c c c c c c c c c c c c c c c c}
						\hline
						Cams & $1$ & $2$ & $3$ & $4$ & $5$ & $6$ & $7$ & $8$ & $9$ & $10$ & $11$ & $12$ & $13$ & $14$ & $15$ & $16$ \\
						\hline \hline
						Mode$-1$ (high amplitude) & $5$ & $4.9$ & $4.7$ & $4.3$ & $3.9$ & $3.3$ & $2.6$ & $1.9$ & $1.2$ & $0.5$ & $-0.2$ & $-0.6$ & $-1.2$ & $-1.6$ & $-1.9$ & $-2$ \\
						Mode$-1$ (low amplitude)~ & $2.5$ & $2.4$ & $2.3$ & $2.1$ & $1.9$ & $1.6$ & $1.3$ & $0.9$ & $0.6$ & $0.2$ & $-0.1$ & $-0.3$ & $-0.6$ & $-0.8$ & $-0.9$ & $-1$ \\
						Mode$-2$ & $5$ & $4.7$ & $4$ & $2.9$ & $1.6$ & $0.3$ & $-0.8$ & $-1.6$ & $-2$ & $-1.9$ & $-1.5$ & $-0.9$ & $-0.1$ & $0.6$ & $1.2$ & $1.5$ \\
						Mode$-3$ & $5$ & $4.5$ & $3$ & $1.2$ & $-0.6$ & $-1.7$ & $-2$ & $-1.4$ & $-0.4$ & $0.7$ & $1.4$ & $1.5$ & $0.9$ & $0$ & $-0.8$ & $-1.2$ \\
						\hline
					\end{tabular}
					\caption{Amplitudes (in mm) of the different cams of the generator in the different mode profiles we used. The first cam is located at $r=0$.}
					\label{tab:shafts}
				\end{center}
			\end{table}
			
				Experiments were conducted in a cylindrical plexiglas tank of the same diameter as the generator to respect the boundary condition~\eqref{eq:IW12}. This transparent cylindrical tank was set into a square plexiglas tank to prevent the experiment visualisation suffering from optical deformations that would occur due to the curved interface created by the cylinder. Both tanks were filled with salt-stratified water with the same density profile. We used the double-bucket method to fill the tanks with a linear stratification~\cite{fortuin1960, oster1963}. Density and buoyancy as a function of depth were measured using a calibrated PME conductivity and temperature probe mounted on a motorized vertical axis. Buoyancy frequency is estimated from the mean value of the $N$ profile obtained from the density function $\rho(z)$. Errors on the buoyancy frequency are estimated using the standard deviation of this $N$ profile, and are in most cases about $4\%$ of the estimated $N$ value. We used buoyancy frequencies in the range $N~\approx~0.6\mathrm{~rad\cdot s^{-1}}$ to $N~\approx~1\mathrm{~rad\cdot s^{-1}}$.
		
		Velocity fields were obtained via Particle Image Velocimetry (PIV). A laser sheet was created by a laser beam (Ti:Sapphire, $2\mathrm{~watts}$, wavelength $532\mathrm{~nm}$) going through a cylindrical lens. It could be oriented either horizontally (to measure the radial and orthoradial velocity) or vertically (to measure the vertical and the radial velocity). For the purpose of visualisation, $10\mathrm{~\mu m}$ diameter hollow glass spheres of volumetric mass $1.1\mathrm{~kg\cdot L^{-1}}$ were added to the fluid while filling the tank. To get good quality velocity fields at the bottom of the tank and while imaging in a horizontal plane, $10\mathrm{~\mu m}$ silver-covered spheres of volumetric mass $1.4\mathrm{~kg\cdot L^{-1}}$ were added when needed in some experiments. Images were recorded at $1\mathrm{~Hz}$ and data processing of the PIV raw images was done using the CIVx algorithm~\cite{fincham2000}.
	
	\section{Results} 
	
		\subsection{Radial Modes}
					
\begin{figure}[!htbp]
				\begin{center}
					\begin{tabular}{>{\centering\arraybackslash}p{0.225\textwidth} >{\centering\arraybackslash}p{0.225\textwidth} >{\centering\arraybackslash}p{0.225\textwidth} >{\centering\arraybackslash}p{0.225\textwidth}}
						\hline 
						$(1)$ Mode & $(2)$ $v_z$ (vertical) & $(3)$ $v_r$ (vertical) & $(4)$ $v_r$ (horizontal)  \\ 
						\hline \hline
						\multicolumn{4}{c}{\epsfig{file=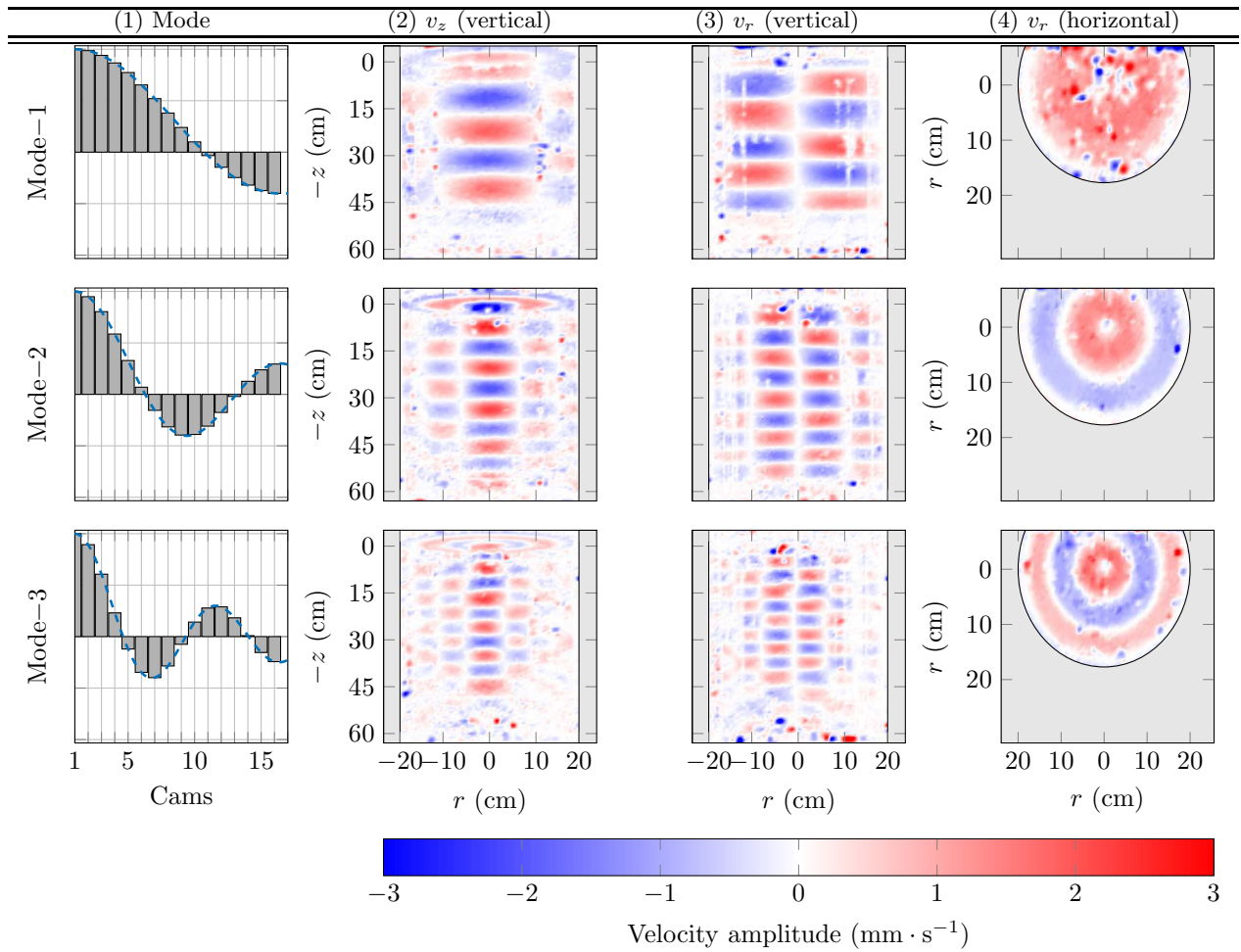}} \\ 
						\hline 
					\end{tabular}
					\caption{Radial modes $1$, $2$, and $3$, as observed in PIV in the experiment. First column: generator configuration that sets the mode. Second column: vertical velocity in a vertical plane. Third and fourth columns: radial velocity in a vertical and in a horizontal plane. Shaded areas are outside of the confining cylinder. For the purpose of visualisation, negative values of $r$ are used in the vertical PIV plane, leading to antisymmetric radial velocities as predicted by equation~\eqref{eq:IW41}.}
				\label{fig:fig3}
				\end{center}
			\end{figure}

			Figure~\ref{fig:fig3} presents a summary of the experimental PIV results for the generation of modes $1$ through $3$ in a linear stratification with $\omega/N = 0.6$ for modes $1$ and $2$ and $\omega/N = 0.65$ for mode$-3$, and a generator amplitude $a = 5\mathrm{~mm}$. The generator plate configuration for each mode is illustrated in the left hand column, with $n$ nodes for mode$-n$. The vertical cross sectional plots of the vertical velocity, presented in column $2$, possess the horizontal structure of the generator, increasingly intricate for the higher modes, with associated vertical sequences of maxima and minima. Columns $3$ and $4$ in figure~\ref{fig:fig3} present vertical and horizontal cross sectional plots of the radial velocity component. For every mode, the radial velocity structure possesses a left-right antisymetry in the vertical plane. The different nodes of radial velocity, which correspond to anti-nodes of vertical velocity, are also clearly visible in plots of the velocity in the horizontal plane, presented in column $4$; these images also show the form of the generator being reproduced by the underlying wave field. No orthoradial velocity $v_\theta$ was observed in the horizontal plane.
			
\begin{figure}[!htbp]
				\centering
				\epsfig{file=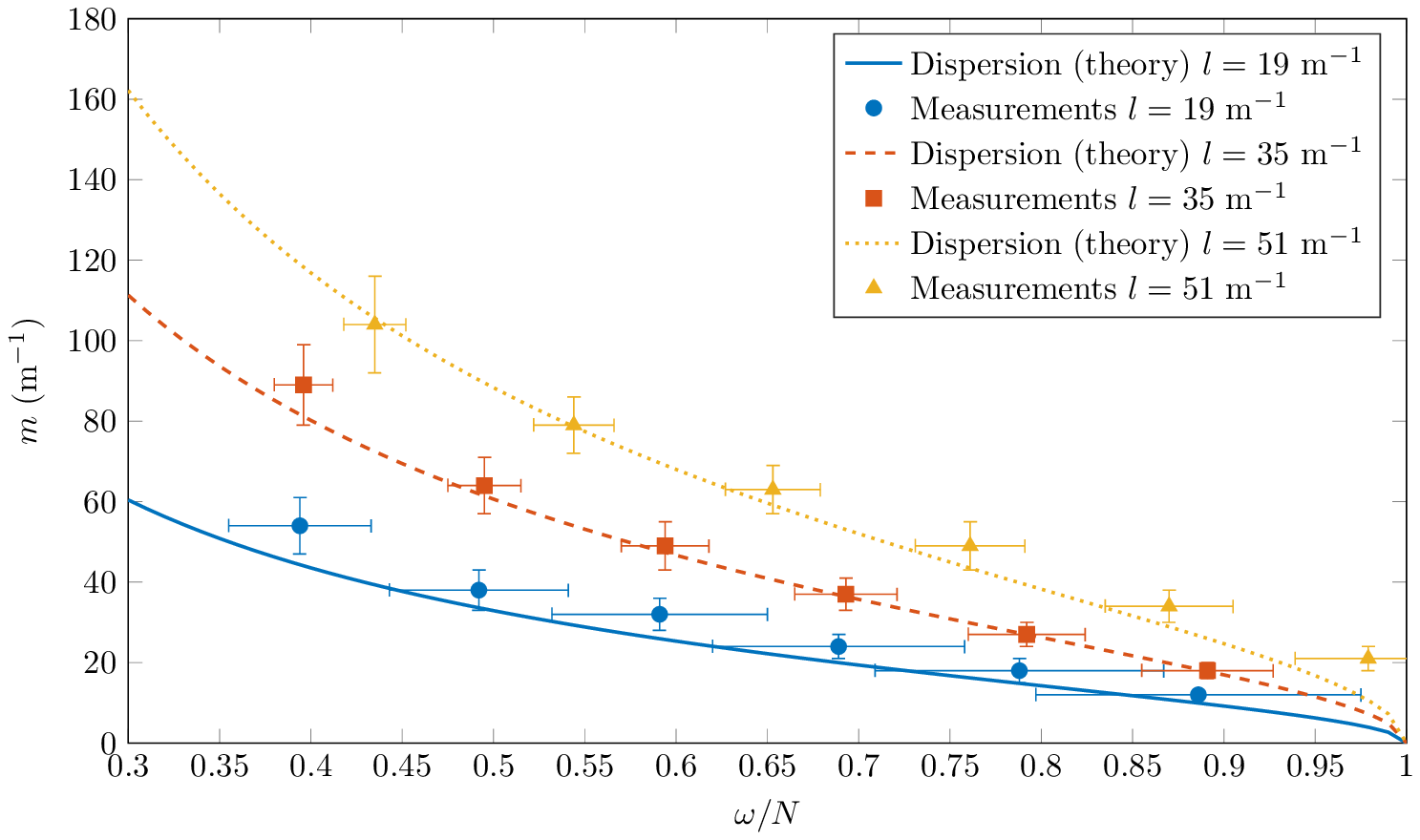}
				\caption{Measured values of the vertical wave number (data points) for modes $1$, $2$, and $3$, compared to the theoretical expectations from equation~\eqref{eq:IW10}.}
				\label{fig:fig4}
			\end{figure}

			In a previous study, Maurer \textit{et al}~\cite{maurer2017} experimentally measured the internal wave dispersion relation for freely propagating waves generated by an axisymmetric wave generator, which was consistent with theoretical predictions. In the modal configuration, however, the dispersion relation does not explicitly contain an angle of propagation, only a statement of the vertical wavelength as a function of the forcing frequency and horizontal wave number. The vertical wave number $m$ was measured for different frequencies $\omega/N$ for the three modes in our experiments. Figure~\ref{fig:fig4} compares the experimental values of $m$ with the theoretical one from equation~\eqref{eq:IW10}, when $f=0$. Measurements were performed by looking at the spatial vertical period of the vertical velocity on PIV images. It shows a good agreement for the three modes considered in the study, though there is a slight deviation at low frequencies, probably because of the error on $N$ which was about $10\%$ for mode$-1$ experiment and $4\%$ for mode$-2$ and mode$-3$ experiments.
			
\begin{figure}[!htbp]
				\centering
				\epsfig{file=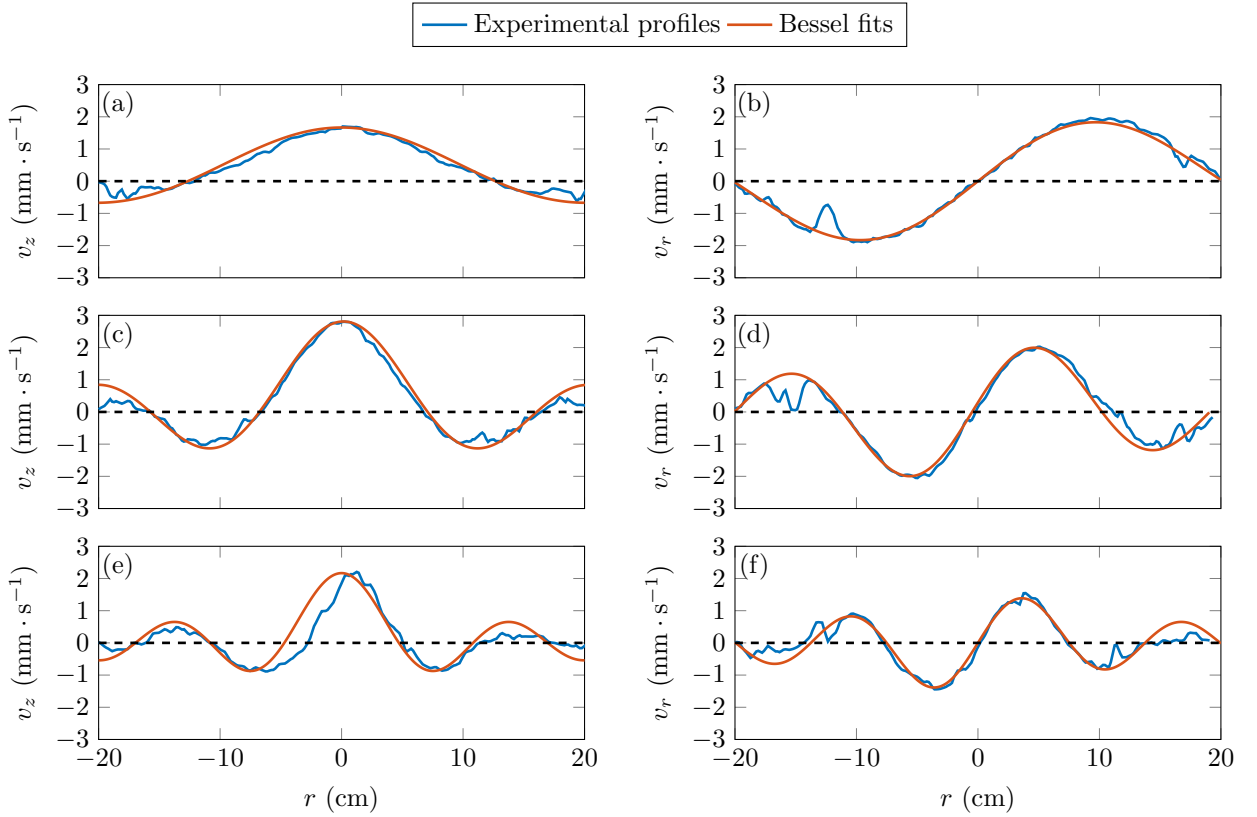}
				\caption{From top to bottom: examples of experimental velocity profiles (left) $v_z$ and (right) $v_r$ for mode$-1$, mode$-2$, and mode$-3$, taken at a given time and altitude, fitted by appropriate Bessel functions: $J_0(lr)$ for $v_z$ and $J_1(lr)$ for $v_r$, with $l=19\mathrm{~m^{-1}}$, $l=35\mathrm{~m^{-1}}$, and $l=51\mathrm{~m^{-1}}$ respectively.}
				\label{fig:fig5}
			\end{figure}

			To quantitatively investigate how close the experiments reproduce the theoretical modal Bessel profile, figure~\ref{fig:fig5} presents radial profiles of $v_z$ and $v_r$, fitted to the expected radial dependency of the Bessel mode, for mode$-1$, mode$-2$, and mode$-3$ configurations. We see that $v_z(r) \propto J_0(l r)$ and $v_r(r) \propto J_1 (l r)$, with $l=19\mathrm{~m^{-1}}$, $35\mathrm{~m^{-1}}$, or $51\mathrm{~m^{-1}}$, as expected; these horizontal structures are preserved through the vertical propagation of the wave field. Small deformations sometimes appear close to the boundaries at $r=20\mathrm{~cm}$, due to boundary layer effects. The perturbation observed symetrically around $12\mathrm{~cm}< |r| < 16\mathrm{~cm}$ is actually caused by laser reflections in the cylinder, producing locally poor PIV visualisation. 
			
		\subsection{Generator Efficiency}
		
		In order to investigate the resonance phenomena that is the topic of the following section, it is essential to reliably measure the amplitude of the internal wave field. Measuring the amplitude of internal waves in closed domains is a delicate task, however, more challenging than measuring their frequency or wavelength, due to unavoidable reflections. As such, we sought a robust measurement of wave amplitude and, further, to demonstrate it through studies of the efficiency of the wave generator, this being the ratio of the amplitude of the waves produced to the amplitude of the generator motion

		A few previous studies have made direct measurements of velocity amplitude, although these are typically done either at high frequencies or relatively high amplitudes. Mathur \& Peacock~\cite{mathur2010} studied transmission and reflection of internal wave beams across a transmission region and took a Fourier transform of the reflected and transmitted wave fields along appropriately chosen transects, Maurer~\cite{maurerPhD} measured wave amplitudes by looking at the maximum of the velocity over a given spatial area, and Supekar~\cite{supekar2018} utilized the distribution of maxima of amplitudes for a velocity field in a widespread two dimensional beam. In performing our experiments, it was necessary to more rigorously define our amplitude measurement methodology based on understanding of the group velocity of the wave fields we were studying.

	The procedure to determine the wave amplitude was the following. In a first step, experimental amplitudes at a given time $t_m$ were determined by fitting a Bessel function to the instantaneous horizontal profile at a given depth $z_{m}$ of the vertical velocity, as illustrated in figure~\ref{fig:fig5}. The depth $z_{m}$ chosen for this profile was selected to be $15\mathrm{~cm}$ below the generator, as the wave field was properly developed at this depth. Since the stratification, the forcing frequency and the radial wavelength are imposed, the only free parameter for the fit is the amplitude of the Bessel function. Note that we used the vertical velocity field for this fitting, since it has larger amplitudes than the radial velocity profile (which is characterized by a node at $r=0$) and so was more amenable to fitting. 
					
\begin{figure}[!htbp]
				\centering
				\epsfig{file=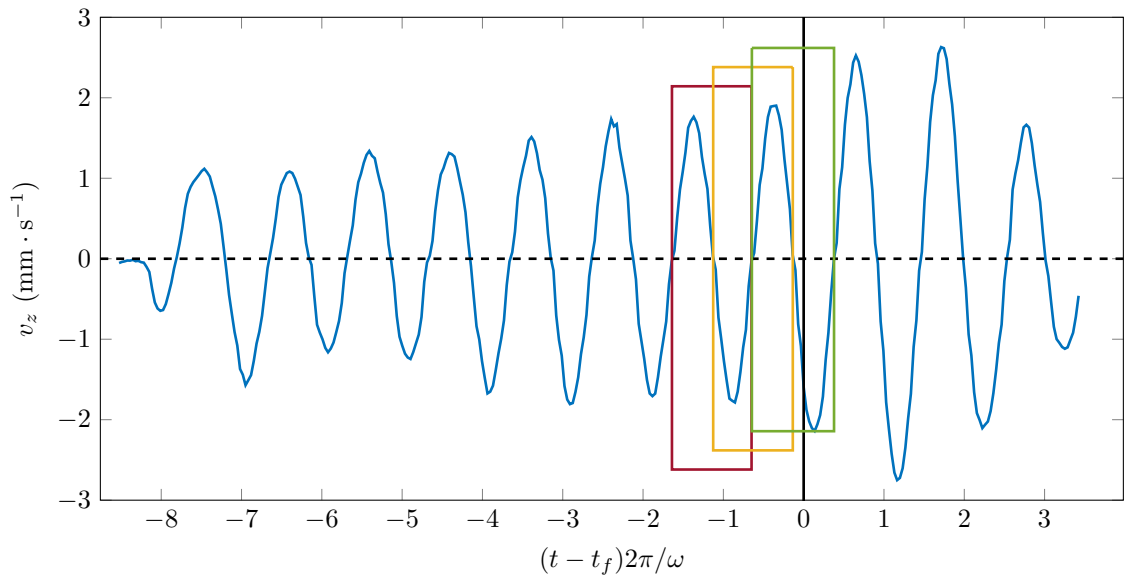}
				\caption{Example of time series of fitted instantaneous Bessel function amplitudes of the vertical velocity, measured at $z=z_m=-15\mathrm{~cm}$ for a mode$-1$ wave. After identifying $t_f$ (solid vertical line), three periods half-covering each other (rectangles) are used to extract the global wave amplitude \textit{via} RMS estimates.}
				\label{fig:fig6}
			\end{figure}

			Measurements were repeated for all images over a time interval $t_m\in [t_i;~t_f]$, with $t_i$ being the time when the wave is expected to first cross the horizontal cross-section at $z=z_{m}$. The time $t_f$ is the time when the reflected wave is predicted to return from the bottom of the tank and reach $z=z_{m}$, resulting in a disturbance of the wave field. Both $t_i$ and $t_f$ were estimated using the group velocity of the wave field established in equation~\eqref{eq:IW14b}. This series of measurements provided a time-series of local wave amplitudes at $z=z_{m}$. An example of such a time series is shown in figure~\ref{fig:fig6}. One would expect this time signal to be sinusoidal. As can be seen in this example, the growth of the wave amplitude can still be slightly observed in the first periods, and the disturbance due to the returning wave is also observed for positive times (after $t=t_{f}$). This illustrates the difficulty of wave amplitude measurement in a finite size tank.
			
			In a second step, in order to best estimate the wave amplitude of the steady state before the reflected wave returned (there is some uncertainty on the exact return time), we computed the RMS value of the time signal over three periods close to $t_f$ half-covering each other, the middle one being just before theoretically seeing the reflected wave, the previous period covering the first half of this one, and the following period covering the second half of it (these measurement windows are illustrated by 3 rectangles in figure~\ref{fig:fig6}). The experimental global amplitude was determined as the mean value of the 3 RMS values obtained (multiplied by $\sqrt{2}$), and the standard deviation of these 3 measurements gives an estimate of the associated error. We checked that the method was sound by repeating some test measurements for other horizontal planes and obtaining consistent results.
			
\begin{figure}[!htbp]
				\centering
				\epsfig{file=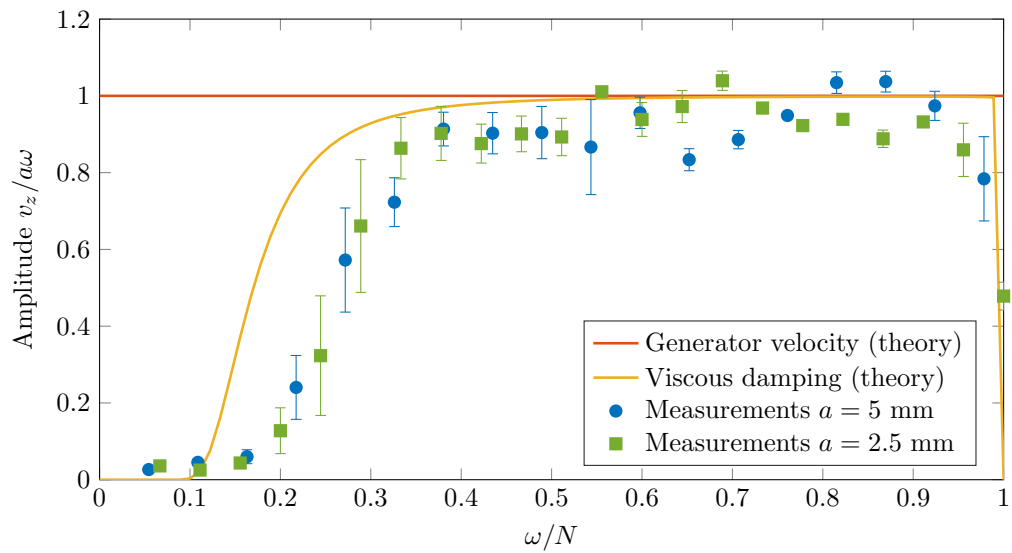}
				\caption{Generator efficiency measured at $z=-15\mathrm{~cm}$, for two experiments with $a=5\mathrm{~mm}$ and $a=2.5\mathrm{~mm}$. Results are compared to the theoretical predictions in the inviscid case and in the weakly viscous case (viscous damping curve, computed for $N=0.9\mathrm{~rad\cdot s^{-1}}$).}
				\label{fig:fig7}
			\end{figure}

			The results of our efficiency experiments are presented in figure~\ref{fig:fig7} for two different generator amplitudes. We plot the velocity amplitude normalised by the generator velocity amplitude $a\omega$. From equation~\eqref{eq:IW26}, without dissipation effects, one expects this ratio to be 1 (straight line in figure~\ref{fig:fig7}). This proves correct in the high frequency range ($0.5<\omega/N < 0.9$), except close to the buoyancy frequency as discussed below. In contrast, at low frequency, the efficiency decreases. This decrease can be interpreted by viscous effects. Indeed, when one includes viscous dissipation in the theoretical development, the expression of the stream function is given by equation~\eqref{eq:IW40}. The curve corresponding to the vertical velocity extracted from this equation (at a depth of $-15\mathrm{~cm}$ since equation~\eqref{eq:IW40} depends on $z$) is plotted in figure~\ref{fig:fig7}, showing a fair agreement with the experimental data points for the two forcing amplitudes $a=2.5 \mathrm{~mm}$ and $a=5 \mathrm{~mm}$. At very low frequencies (below $\omega/N = 0.05$), the amplitude is so low that measurements become impossible. Finally, for $\omega/N = 0.9$ to $1$, we notice a decrease in amplitude that is expected, as shown by the theoretical curve, since the waves are evanescent for $\omega/N > 1$. However, this decrease comes sooner than expected. 
			 
			 To conclude, the generator efficiency was investigated and shows a similar behaviour as the theoretical prediction in all frequency ranges, providing one takes viscous effects into account. In addition, we identify a range of frequencies, from $\omega/N=0.5$ to $\omega/N=0.9$, where there is a very strong agreement with the non viscous theory, making this range suitable for axisymmetric modes experiment and for resonant enhancement.
		
		\subsection{Resonance}
		
		Having established the response of the stratification to the wave generator forcing, we then conducted experiments to detect resonance for a mode$-1$ excitation, due to the multiple reflections of the wave field at the top and bottom boundaries of the tank. These experiments consisted of measuring the amplitude of the wave after a time long enough to ensure the establishment of the steady state resonant wave field. In order to allow a minimum of about $10$ back and forth crossings, we chose this time to be $280\mathrm{~s}$, based of the minimum value of the group velocity of the waves. We thus looked for the maximum of vertical velocity at $r=0$ in a time period between $280$ to $300\mathrm{~s}$ (end of experiment duration) after the generator forcing was initiated. Using the vertical velocity at $r=0$ is optimal, as it is the highest velocity we can measure in the whole tank, and is straightforward since it corresponds to the amplitude of the velocity field.
				
\begin{figure}[!htbp]
				\centering
				\epsfig{file=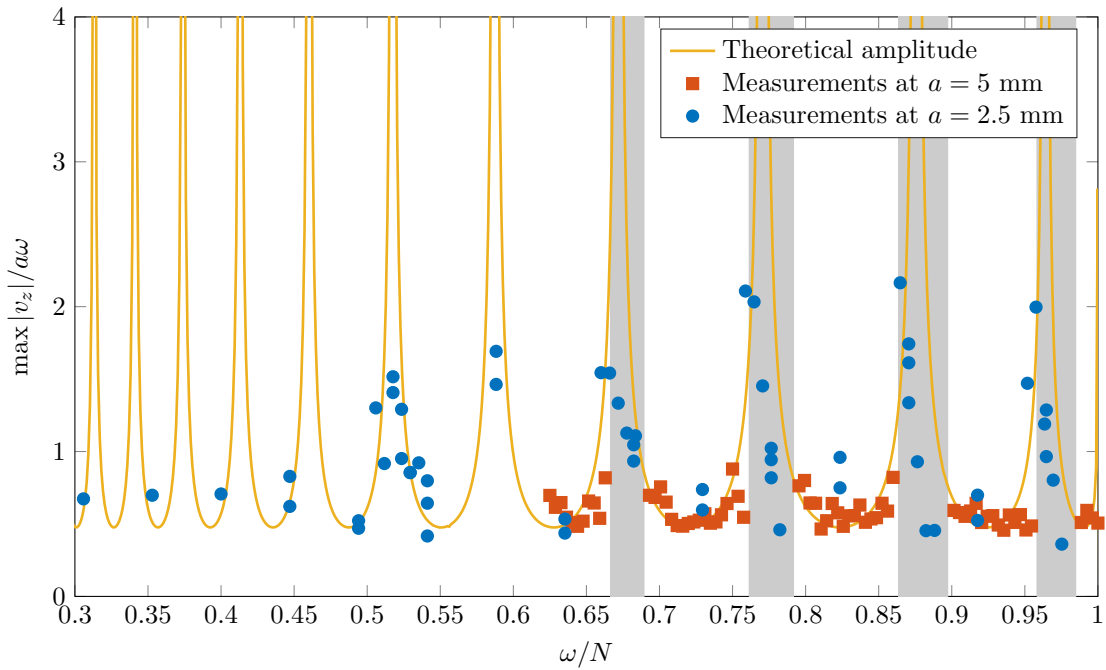}
				\caption{Amplitude measurements of the vertical velocity in the resonant cavity normalised by the generator velocity amplitude. Yellow line: theoretical amplitude for an infinite sum of waves as a function of the frequency $\omega/N$. The dots are from three different sets of measurements run for different buoyancies or amplitudes. Light gray regions show interval of frequencies in which non-linear effects are clearly seen in the experiment for $a=5\mathrm{~mm}$.}
				\label{fig:fig8}
			\end{figure}

		Our experimental results are presented in figure~\ref{fig:fig8}, showing the measured velocity amplitude, normalised by the generator velocity amplitude $a\omega$. We performed two sets of experiments : one with $a=2.5\mathrm{~mm}$ and $N\simeq 0.90 \mathrm{~rad \cdot s^{-1}}$ (blue circles) and another one with $a = 5\mathrm{~mm}$ and $N\simeq 0.88 \mathrm{~rad \cdot s^{-1}}$ (red squares). The first set of experiments was mainly aimed at identifying the resonant peaks; the second set was more evenly spread over all frequencies (a hundred values of $\omega/N$ from $0.625$ to $1$ at a regular interval). In the latter case, however, because of the larger generator amplitude, all experiments where the frequency was too close to the resonance led to strong non-linear effects, making the measurement of an amplitude impossible. For this reason, the corresponding data points are not shown. The theoretical curve for the maximal amplitude of vertical velocity normalised by the generator, computed from equation~\eqref{eq:IW35}, is also plotted in figure~\ref{fig:fig8} as a solid line.
		
		With the generator configured at low forcing amplitude ($a=2.5\mathrm{~mm}$), the peaks corresponding to the first resonant frequencies were observed as predicted by the theory (see table~\ref{tab:resonantomegas}). The measured resonance peaks are not exactly centered on the predicted resonant frequencies, but this is not inconsistent with the characteristic $4\%$ error on $N$. We see that in the vicinity of resonant frequencies the wave field reaches twice the amplitude of the generator, and even more for the highest frequencies. For non-resonant frequencies, however, the wave interaction is destructive and the measured amplitude is half the amplitude of the generator.
			
\begin{figure}[!htbp]
				\centering
				\epsfig{file=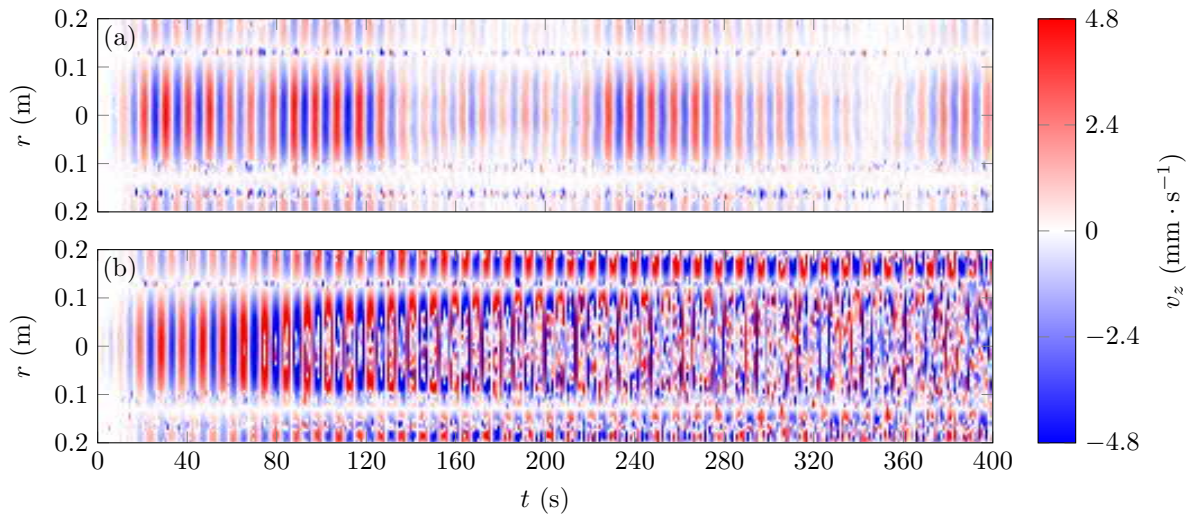}
				\caption{Temporal evolution of a horizontal profile of vertical velocity located at the center of the tank, for (a) $\omega/N = 0.73$ and (b) $\omega/N = 0.77$. These profiles are measured at mid-depth in the tank for a mode$-1$ excitation, with $a=5\mathrm{~mm}$.}
				\label{fig:fig9}
			\end{figure}

			In the vicinity of a resonant excitation frequency, we observed that wave field amplitude kept strengthening until it triggered substantial non-linear effects. To illustrate this, figure~\ref{fig:fig9}  presents the temporal evolution of a horizontal profile of the vertical velocity component for $\omega/N=0.73$ (figure~\ref{fig:fig9}(a), non-resonant), and $\omega/N = 0.77$ (figure~\ref{fig:fig9}(b), resonant). In the non-resonant case, each velocity profile has the shape of a Bessel profile, which is conserved during the whole experiment. No non-linear deformation of the wave field can be observed. The amplitude shows a beating behaviour, because the reflected wave fields are not opposite in phase, due to the non-resonance condition. In the resonant case, such oscillations do not exist as all reflections are in phase. These reflections, however, lead to an increasing amplitude that quickly triggers non-linear effects, after $80$ seconds in the example in figure~\ref{fig:fig9}(b). The Bessel axisymmetric shape of the profile starts to disappear from the center of the tank due to emerging non-linear features, as the amplitude is maximum at $r=0$. The non-linearities then propagate radially towards the boundaries of the cylindrical tank, and the velocity field does not have a modal shape anymore (after $250$ seconds in figure~\ref{fig:fig9}(b)). 
			
			By performing a similar analysis for all frequencies in the large amplitude case ($a=5\mathrm{~mm}$), we identified four frequency ranges in which all experiments led to non-linear effects. These ranges are marked with grey zones in figure~\ref{fig:fig8}. These intervals show a good agreement with the predicted resonance peaks (table~\ref{tab:resonantomegas}) and with the increasing amplitude observed for the low amplitude measurements ($a = 2.5\mathrm{~mm}$). 			 
		
	\section{Conclusions and Discussion}

		We have presented the results of a laboratory experimental study of axisymmetric internal wave mode generation, incorporating both  radial and vertical confinement. To  support the experiments, we developed the theoretical framework of radial modes propagating vertically in uniform stratifications, accounting for the impact of weak viscous damping. The effect of rotation was not explored in our experiments, but the governing equations predict qualitatively similar behaviour as in the non-rotating case, the impact of rotation being foremost to influence the vertical wavenumber of the wave field for a given forcing frequency and buoyancy frequency. The experimental wave fields were produced using a novel configuration of internal wave generator technology that has previously been primarily used to excite nominally planar wave fields; in our experiments the arrangement directly excited the Bessel functions that are the natural basis of cylindrical modes.
		
	For the basic structure of the wave fields, there was very good qualitative and quantitative agreement between experiments and theory. Modes $1$ through $3$ were excited, leading to vertical and radial velocity profiles consistent with associated Bessel function forcing, and confirming the expected dispersion relation. As an additional component of these studies, we determined the efficiency of modal excitation by carefully studying the fluid system response to the generator forcing, fitting the PIV data to Bessel functions. A range of frequencies, from $\omega/N = 0.5$ to $0.9$, was identified as being particularly suitable for studying axisymmetric modes as in this frequency range the wave field is attenuated very little in the tank and has an almost full response to the forcing amplitude of the generator.
	
	Having established the ability to excite vertical modes, the role of vertical confinement was then investigated. Such confinement has the potential to generate a resonance effect when reflected modes constructively interfere with each other. The resonance conditions for our system were determined and a series of experiments with different forcing amplitudes were performed. The experimental results of wave field amplification aligned well with resonance predictions. Within the bounds of resonant peaks, the wave field was seen to amplify sufficiently to trigger nonlinear effects that then eroded the linear wave field structure outwards from the centerline of the experimental domain, ultimately leading to a fully non-linear wave field throughout the experimental domain.
		
	While there have been a number of nominally two-dimensional experimental studies comparing plane wave or mode behaviour with theoretical models, considering both their spatiotemporal form and transition to nonlinear phenomena, there have been few such studies for axisymmetric geometries, which have typically been limited to studying the wave field excited by an oscillating sphere. Axisymmetric wave fields are arguably more relevant as fundamental configurations for studying scenarios such as the excitation of atmospheric internal wave fields by storm cells~\cite{alexander1995} and the excitation of near-inertial wave fields in the ocean by surface storms~\cite{alford2003}. The experimental apparatus and consequent studies presented here demonstrate a new ability to excite axisymmetric wave fields and pure radial modes, opening the path to investigation of linear (e.g. internal wave transmission) and nonlinear (e.g. TRI) internal wave phenomena in axisymmetric geometries. For example, a recent study by Duran-Matute \textit{et al}~\cite{duranmatute2013} has used focusing by an oscillating torus to study the onset of instabilities; the vertical mode geometry we have developed and demonstrated here provides an arguably even more canonical configuration in which to study such non-linear phenomena, since any axisymmetric wave field can be expressed in terms of Bessel functions.
	
\medskip
\textbf{Acknowledgements:}

This work has been partially supported by by the ANR through grant ANR-17-CE30-0003 (DisET) and by ONR Physical Oceanography Grant N000141612450.


\end{document}